\documentclass[aps,prl,preprint,superscriptaddress]{revtex4}

\usepackage{graphicx}
\usepackage{dcolumn}
\usepackage{bm}

\begin{document}


\title{Optical Implementation of Non-locality with Coherent
Light Fields for Quantum Communication}


\author{Kim Fook Lee}
\email[]{kflee@mtu.edu}
\affiliation{%
Department of Physics,\\ Michigan Technological
University,\\ Houghton, Michigan 49931}%


\date{\today}

\begin{abstract}

Polarization correlations of two distant observers are observed by
using coherent light fields based on Stapp's formulation of
nonlocality. Using a 50/50 beam splitter transformation, a
vertically polarized coherent light field is found to be entangled
with a horizontally polarized coherent noise field. The superposed
light fields at each output port of the beam splitter are sent to
two distant observers, where the fields are interfered and
manipulated at each observer by using a quarter wave plate and an
analyzer. The interference signal contains information of the
projection angle of the analyzer, which is hidden by the phase
noises. The nonlocal correlations between the projection angles of
two distant observers are established by analyzing their data
through analog signal multiplication without any post-selection
technique. This scheme can be used to implement Ekert's protocol for
quantum key distribution.

\end{abstract}


\maketitle


Entanglement and superposition are foundations for the emerging
field of quantum communication and information processing.
Generally, implementation of an optical quantum information system
is based on two types of quantum variables; discrete variable and
continuous variable. They are usually generated through nonlinear
interaction process in $\chi^{(2)}$~\cite{Kwiat95} and
$\chi^{(3)}$~\cite{kim06,kim08} media. Discrete-variable qubit based
implementations using
polarization~\cite{liang06,chen07,Jun08,chuang07,sharping06} and
time-bin~\cite{Brendel99,Tittel98,Tittel99} entanglement are
difficult to obtain unconditional-ness and usually have low optical
data-rate because of post-selection technique with low probability
of success in a single photon
detector~\cite{liang06,chuang07,Liang05}. Continuous-variable
implementations using quadrature
entanglement~\cite{Furusawa04,Ralph03,Leuch02} and polarization
squeezing~\cite{Ralph02} could have high efficiency and high optical
data-rate because of available high speed and efficient homodyne
detection, and hence usually obtain unconditional-ness. However, the
quality of quadrature entanglement is very much depended on the
amount of squeezing which is very sensitive to loss, so the
quadrature entanglement is imperfect for implementing any
entanglement based quantum protocols over long distance.
Continuous-variable protocols which are not based on entanglement,
for instance, coherent-state based quantum key
distribution~\cite{Yuen04,Corndorf03,Barbosa03,
Grangier02,Grangier03, Dowling08}, is perfect for long distance
quantum cryptography.

Entanglement distribution over long distance is an important
experimental challenge in quantum information processing because of
unavoidable transmission loss associated with low coupling
efficiency from free space to optical fibers. There are few
experimental approaches to resolve loss tolerant by using coherent
light source. Optical wave mechanics
implementations~\cite{kim05,kim02} of entanglement and superposition
with coherent fields (coherent state with large mean photon numbers)
have been demonstrated. This implementation has been used to study
entanglement swapping and tests of non-locality~\cite{kim02}. In the
similar approach, coherent fields have played an important role in
quantum computing such as search
algorithm~\cite{lloyd00,Bhattacharya02} and factorization of
numbers~\cite{Schleich08}. Optical wave illustration of quantum
phenomena such as negative valued of Wigner function for transverse
position ($X_\perp$) and transverse momentum or angle ($P_\perp$) of
a coherent light field has been performed~\cite{kim99}.

In this paper, two orthogonal coherent light fields with mean photon
number around $10^7$ per unit bandwidth are used to implement
Stapp's formulation of two distant observers~\cite{Stapp99}.
Electric field fluctuations of these two light fields are
negligible. In order to achieve randomness in phase fluctuations,
one of the coherent light fields is modulated with a pseudo-random
noise generator. To understand the essence of this work, a brief
description of Stapp's formulation for nonlocal correlation function
(expectation value) of two distant observers is discussed.

\begin{figure}
\centering
\includegraphics[scale=0.35]{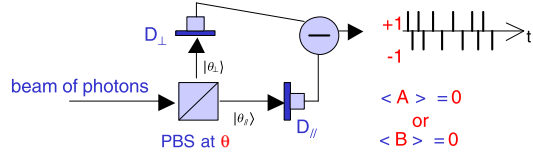}
\caption{\label{fig:1}Detection scheme based on balanced homodyne
detection for measuring operators $A_{1}$ and $B_{2}$.}
\end{figure}

In the Stapp's approach~\cite{Stapp99} for a two photon entangled
state
$|\psi^{\pm}\rangle$=$\frac{1}{\sqrt{2}}\left(|H_{1}V_{2}\rangle\pm
|V_{1}H_{2}\rangle\right)$, the two entangled photons are sent to
two spatially separated measuring devices A and B. Device A is an
analyzer for projecting the linear polarization of the incoming
photon. When the analyzer A is oriented along the polarization angle
$\theta_{1}$, the polarization state of the incoming photon is
projected onto the state,
\begin{eqnarray}
|\theta_{1}\rangle&=&\cos\theta_{1}|H_{1}\rangle +
\sin\theta_{1}|V_{1}\rangle \label{eq:noise1}
\end{eqnarray}
where H and V are horizontal and vertical axes. The corresponding
orthogonal polarization state is given by
\begin{eqnarray}
|\theta_{1}^{\perp}\rangle&=&-\sin\theta_{1}|H_{1}\rangle +
\cos\theta_{1}|V_{1}\rangle \label{eq:noise2}
\end{eqnarray}
The operator associated with analyzer A can be represented as
$A_{1}$, which is defined as~\cite{Stapp99}
\begin{eqnarray}
A_{1}&=&2|\theta_{1}\rangle \langle
\theta_{1}|-\left(|\theta_{1}\rangle \langle
\theta_{1}|+|\theta_{1}^{\perp}\rangle \langle
\theta_{1}^{\perp}|\right)\,. \label{eq:noise3}
\end{eqnarray}
The operator $A_{1}$ has eigenvalues of $\pm 1$, such that,
\begin{eqnarray}
A_{1}|\theta_{1}\rangle &=& 1\,|\theta_{1}\rangle\nonumber\\
A_{1}|\theta_{1}^{\perp}\rangle &=& -1\,|\theta_{1}^{\perp}\rangle\,
\label{eq:noise4}
\end{eqnarray}
depending on whether the photon is transmitted ($\parallel$) or
rejected ($\perp$ ) by the analyzer. Similarly, the analyzer B
oriented along polarization angle $\theta_{2} $ can be defined as
operator $B_{2}$. One should note that the operator $A_1(B_2)$ with
eigenvalues of $\pm 1$ could be measured by using the detection
scheme as shown in Fig.~\ref{fig:1}. Two detectors are placed at the
two output ports of a cube polarization beam splitter (PBS). Their
output currents are subtracted from each other. The arrangement of
this detection scheme can be used for measuring operator $A_1$ of
Eq.(3), that is the subtraction between the projection
 of the transmitted signal $|\theta_{1}\rangle_{\parallel}\langle
\theta_{1}|$ and the projection of the reflected signal
$|\theta_{1}\rangle_{\perp}\langle \theta_{1}|$. Let's consider a
beam of photons incidents on the PBS, if one photon goes through the
PBS, it will produce non-zero signal at detector $D_{\parallel}$ and
zero signals at detector $D_{\perp}$. Then, the subtraction yields
positive signal as of $D_{\parallel}-D_{\perp}\geq 0$. If a photon
is reflected from the PBS, it will go to the detector $D_{\perp}$
and produce non-zero signal at detector $D_{\perp}$ and zero signals
at detector $D_{\parallel}$. Then, the subtraction yields negative
signal as of $D_{\parallel}-D_{\perp}\geq 0$. For a certain amount
of time, the subtraction records the random positive and negative
spikes corresponding to the eigenvalues of  +1 and -1 of operator
$A_{1}$, respectively, as shown in the inset of Fig.~\ref{fig:1}.
The incoming photons are in the superposition of
$|\theta_{1}\rangle_{\parallel}$ and $\theta_{1}\rangle_{\perp}$.
Hence, as the time elapses, the detection scheme A records a series
of discrete random values, +1 and -1. Then, for a state with equal
probabilities of $\parallel$ and $\perp$ photons, the mean value of
$A_{1}$ is zero, that is $\langle A_{1}\rangle$ = 0. Similarly, we
can apply the same detection scheme for measuring operator $B_{2}$.
We will obtain $\langle B_{2}\rangle = 0$. It is very important to
show that the detection scheme exhibits wave-particle duality
principle. The wave character of the operator $A_{1} (B_2)$ is
recognized as interference of the outcomes of $A_{1}(B_{2})$ due to
the linear superposition of the projected states
$|\theta\rangle_{\parallel}\langle \theta|$ and
$|\theta\rangle_{\perp}\langle \theta|$. The particle character of
the operator $A_{1} (B_{2})$ is the discreteness of random values of
+1 and -1. The product of the operators $A_{1}$ and $B_{2}$ or the
multiplication of their output signals will produce correlation
functions, as given by,
\begin{equation}
C_{q}(\theta_{1},\theta_{2})= \langle \psi^{\pm}|
A_{1}B_{2}|\psi^{\pm} \rangle =-\cos2(\theta_{1}\pm\theta_{2})\,.
\label{eq:noise5}
\end{equation}
Eq.5 is usually referred to as the expectation value for the product of
operators $A_1$ and $B_2$. For the other two Bell states,
$|\varphi^{\pm}\rangle=\frac{1}{\sqrt{2}}\left(|H_{1}H_{2}\rangle\pm
|V_{1}V_{2}\rangle\right)$, the correlation functions are given by,
\begin{equation}
C_{q}(\theta_{1},\theta_{2})= \langle \varphi^{\pm}|
A_{1}B_{2}|\varphi^{\pm} \rangle =\cos2(\theta_{1}\mp\theta_{2})\,.
\label{eq:noise5}
\end{equation}.

\begin{figure}
\centering
\includegraphics[scale=0.4]{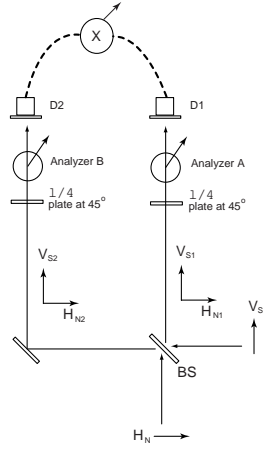}
\caption{\label{fig:2}Experimental setup for demonstrating Stapp's
formulation of nonlocality based on coherent fields.}
\end{figure}

The proof-of principle experimental setup is shown in
Fig.~\ref{fig:2}. The coherent light source is a HeNe laser operated
at 632nm. A vertically polarized beam is a coherent light field $\bf
V_{S}$ with frequency shifted at 110 MHz. A horizontally polarized
beam is a random phase-modulated light field $\bf H_{N}$ induced by
an variable acoustic optics modulator at around 110 MHz, which is
externally added and modulated by a random noise generator. These
two light fields are then combined through a beam splitter. The beam
1 from the output port 1 of the beam splitter contains a
superposition of vertically polarized coherent field and
horizontally polarized noise field. Similarly, for the beam 2 from
the output port 2 of the beam splitter.  A quarter wave plate at
$45^{o}$ as part of measuring device is inserted at beams 1 and 2 to
transform the linearly polarized states to circularly polarized
states. By using a quarter wave plate transformation matric, the
fields amplitudes $\bf V_{S1}$, $\bf H_{N1}$, $\bf V_{S2}$ and $\bf
H_{N2}$ are transformed as,

\begin{eqnarray}
\bf V_{S1}&\rightarrow&-i\hat{\bf H}_{S1}+\hat{\bf V}_{S1}\nonumber\\
\bf H_{N1}&\rightarrow&\hat{\bf H}_{N1}-i\hat{\bf V}_{N1}\nonumber\\
\bf V_{S2}&\rightarrow&-i\hat{\bf H}_{S2}+\hat{\bf V}_{S2}\nonumber\\
\bf H_{N2}&\rightarrow&-\hat{\bf H}_{N2}+i\hat{\bf V}_{N2}\,.\nonumber\\
\label{eq:noise12}
\end{eqnarray}
For simplicity we use unit vector notation and drop the amplitude of
fields notation. Now, analyzer A in beam 1 will experience
homogeneous superposition of left circularly polarized coherent
field and right circularly polarized coherent noise field.
Similarly, for analyzer B in beam 2. Analyzer $A(B)$ is placed
before the detector 1(2) to project out the phase angle
$\theta_{1}$($\theta_{2}$) as,
\begin{eqnarray}
\hat{e}_{1}&=&\cos\theta_{1}\,\hat{H}
+\sin\theta_{1}\,\hat{V}\nonumber\\
\hat{e}_{2}&=&\cos\theta_{2}\,\hat{H} +\sin\theta_{2}\,\hat{V}.
\end{eqnarray}
The superposed field in beam 1 after the $\lambda/4$ plate and the
analyzer is,
\begin{eqnarray}
\textbf{E}_{1}(t)&=&[(\hat{\textbf{H}}_{N1}-i\hat{\textbf{V}}_{N1})e^{-i(\omega+\Omega )t-i\phi}\nonumber\\
&+&(-i\hat{\textbf{H}}_{S1}+\hat{\textbf{V}}_{S1})e^{-i(\omega+\Omega)t}]\cdot \hat{\textbf{e}}_{1}\nonumber\\
&=&(\cos\theta_{1}-i\sin\theta_{1})e^{-i(\omega+\Omega)t-i\phi}\nonumber\\
&+&(-i\cos\theta_{1}+\sin\theta_{1})e^{-i(\omega+\Omega)t}
\label{eq:noise13}
\end{eqnarray}
and similarly for the superposed field in beam 2,
\begin{eqnarray}
\textbf{E}_{2}(t)&=&[(-\hat{\textbf{H}}_{N2}+i\hat{\textbf{V}}_{N2})e^{-i(\omega+\Omega)t-i\phi}\nonumber\\
&+&(-i\hat{\textbf{H}}_{S2}+\hat{\textbf{V}}_{S2})e^{-i(\omega+\Omega)t}]\cdot \hat{\textbf{e}}_{2}\nonumber\\
&=&(-\cos\theta_{2}+i\sin\theta_{2})e^{-i(\omega+\Omega)t-i\phi}\nonumber\\
&+&(-i\cos\theta_{2}+\sin\theta_{2})e^{-i(\omega+\Omega)t}
\label{eq:noise14}
\end{eqnarray}
where $\omega$ and $\Omega$ are optical and modulated frequencies,
and $\phi$ is a random phase of the noise field. Instead of using
the detection scheme as shown in Fig.~\ref{fig:1}, a detector
(Hamamatsu S1223-01 with detection bandwidth of 20 MHz) with a DC
block will provide the similar result except the 3 dB gain in the
balanced detection.

Thus, the interference signals obtained in detectors 1 and 2 can be written
as,
\begin{eqnarray}
D_{1}(\phi)&=&-ie^{i(2\theta_{1}+\phi)}+c.c\nonumber\\
&=&\sin(2\theta_{1}+\phi)\nonumber\\
D_{2}(\phi)&=&ie^{i(2\theta_{2}+\phi)}+c.c\nonumber\\
&=&-\sin(2\theta_{2}+\phi).
\label{eq:noise15}
\end{eqnarray}
The interference signals of Eq.~\ref{eq:noise15} for detectors 1 and
2 are the measurements of operators $A_{1}$ and $B_{2}$,
respectively. The interference signal in detector 2 is
anti-correlated to detector 1 because of the $\pi$ phase shift of
the beam splitter. The interference signals contain information of
the projection angles of the analyzers, which are protected by the
random noise phases, $\phi$. The average of the interference signals
is zero, that is, $\langle A_{1}\rangle$ = 0 and $\langle
B_{2}\rangle$ = 0. To further discuss the significant of measuring
the operator $A_1$, the interference signals obtained in detector 1
can be rewritten as,
\begin{eqnarray}
A_{1}(\phi)&=&\cos(2\theta_{1})\sin(\phi)+\sin(2\theta_{1})\cos(\phi)
\label{eq:noise16}
\end{eqnarray}
Eq.~\ref{eq:noise16} is identical in structure with operator $A_{1}$  as in
Eq.~\ref{eq:noise3}, that is
\begin{eqnarray}
A_{1} &=&\cos 2\theta_{1}(|V_{1}\rangle \langle V_{1}|-|H_{1}\rangle\langle H_{1}|)\nonumber\\
&+&\sin 2\theta_{1}(|V_{1}\rangle \langle H_{1}|+|H_{1}\rangle \langle V_{1}|).\nonumber\\
\label{eq:noise17}
\end{eqnarray}
Note that the unit polarization projectors $(|V_{1}\rangle \langle
V_{1}|-|H_{1}\rangle \langle H_{1}|)$ and $(|V_{1}\rangle \langle
H_{1}|+|H_{1}\rangle \langle V_{1}|)$ in Eq.~\ref{eq:noise17} can be
interpreted by in-phase and out-of-phase or out-of-phase and
in-phase components of the noise field because of random noise
phase, $\phi$. The interference signals in detectors 1 and 2 are
then multiplied to obtain the anti-correlated multiplication signal,
\begin{eqnarray}
A_{1}\times B_{2}&=&-\sin(2\theta_{1}+\phi)\sin(2\theta_{2}+\phi)\nonumber\\
&=&-\cos(2(\theta_{1}-\theta_{2}))-\cos(2(\theta_{1}+\theta_{2}+\phi))\nonumber\\
\label{eq:noise18}
\end{eqnarray}
Then, the mean value of this multiplied signal is measured. We
obtain the correlation function $C(\theta_{1},\theta_{2})$,
\begin{eqnarray}
\overline{A_{1}\times B_{2}}&\propto& C(\theta_{1},\theta_{2})
\propto-\cos(2(\theta_{1}-\theta_{2})) \label{eq:noise19}
\end{eqnarray}
where the random noise phases term in Eq.~\ref{eq:noise18} is
averaging to zero. We have projected out the polarization-entangled
state
$|\psi^{-}\rangle=\frac{1}{\sqrt{2}}[|H_{1}V_{2}\rangle-|V_{1}H_{2}\rangle]$.
We normalized the correlation function $C(\theta_{1},\theta_{2})$
with its maximum obtainable value that is $\theta_{1} = \theta_{2}$.
Thus, for the setting of the analyzers at $\theta_{1} = \theta_{2}$,
the normalized correlation function
$C^{N}(\theta_{1},\theta_{2})=-1$ shows that the two beams are
anti-correlated.

For other Bell's state preparation, such as,
$|\psi^{+}\rangle=\frac{1}{\sqrt{2}}[|H_{1}V_{2}\rangle+|V_{1}H_{2}\rangle]$,
the $\lambda/4$ wave plate at beam 2 is rotated at -$45^{\circ}$,
then the beat signal $B_{2}$ of Eq.~\ref{eq:noise15} is given by
\begin{eqnarray}
B_{2}(\phi)\propto -\sin(2\theta_{2}-\phi)\,.
\end{eqnarray}
Hence, the correlation function of Eq.~\ref{eq:noise19} is
\begin{eqnarray}
C(\theta_{1},\theta_{2})\propto-\cos2(\theta_{1}+\theta_{2})
\end{eqnarray}
corresponding to the projected polarization-entangled state
$|\psi^{+}\rangle$.

As for the state
$|\varphi^{+}\rangle=\frac{1}{\sqrt{2}}[|H_{1}H_{2}\rangle+|V_{1}V_{2}\rangle]$,
a $\lambda/2$ plate in beam 2 is inserted, then the minus sign of
beat signal $B_{2}$ of Eq.~\ref{eq:noise15} is changed to positive
sign. The correlation function of Eq.~\ref{eq:noise19} is $\propto$
$\cos2(\theta_{1}-\theta_{2})$. Thus, the
$C(\theta_{1},\theta_{2})=+1$ for $\theta_{1}=\theta_{2}$, then the
projected polarization-entangled state is perfect correlated that is
$|\varphi^{+}\rangle$.

Similarly, with the $\lambda/2$ wave plate at beam 2 and the
$\lambda/4$ wave plate at beam 2 rotated at -$45^{\circ}$, the beat
signal $B_{2}$ of Eq.~\ref{eq:noise15} is $=\sin(2\theta_{2}-\phi)$.
Thus, the correlation function of Eq.~\ref{eq:noise19} is $\propto$
$\cos2(\theta_{1}+\theta_{2})$ corresponding to the projected
polarization-entangled state
$|\varphi^{-}\rangle=\frac{1}{\sqrt{2}}[|H_{1}H_{2}\rangle-|V_{1}V_{2}\rangle$.
The scheme is perfect for quantum communication processing because
the four Bell states are prepared by just changing the phases in
beam 2. For practical quantum communication, Alice can keep the beam
2 and sent out the beam 1 to Bob. Since Alice can change the phases
of beam 2 locally, her acts will change the non-local correlation
function with Bob.

\begin{figure}
\centering
\includegraphics[scale=0.35]{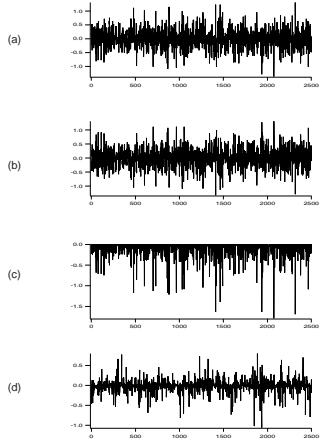}
\caption{\label{fig:3}(a)A single shot of the beat signal at
detector 1 and (b) detector 2; (c) the multiplied beat signal for
$\theta_{1} =\theta_{2}$ and (d) for $\theta_{1}=0,
\theta_{2}=45^{\circ}$.}
\end{figure}

As for an illustration of our experimental observation for the
correlation function
$C(\theta_{1},\theta_{2})=-\cos2(\theta_{1}-\theta_{2})$ of the
state $|\psi^{-}\rangle$, we take a single shot of the
anti-correlated beat signal at detectors 1 and 2 for
$\theta_{1}=\theta_{2}$ as shown in Fig.~\ref{fig:3}a and b
respectively. One may notice that the mean value of beat signal
$\langle A_{1}\rangle$ and $\langle B_{2}\rangle$ are zero as
predicted. The multiplied beat signal is shown in Fig.~\ref{fig:3}c
which has the maximum obtainable mean value. Also shown in
Fig.~\ref{fig:3}d is the multiplied beat signal for the case
$\theta_{1}=0$ and $\theta_{2}=45$, where its mean value
approximately zero as predicted by $C(\theta_{1},\theta_{2})$.

To further verify the nonlocality between these two spatially
separated beams, the correlation functions between two distant
observers ($A_{1}$ and $B_{2})$ are measured for the violation of
Bell Inequality~\cite{Bell78}, which is given by~\cite{Peres93},

\begin{eqnarray}
|C^{N}(a,b)-C^{N}(a,c)|\leq1+C^{N}(b,c)\nonumber\\
\end{eqnarray}
or,
\begin{eqnarray}
F(a,b,c)&=&|C^{N}(a,b)-C^{N}(a,c)|-1- C^{N}(b,c)|\leq 0.\nonumber\\
\label{eq:noise20}
\end{eqnarray}
where $\bf a$, $\bf b$ and $\bf c$ are projection angles of the
analyzers A and B. For the entangled state,
$|\psi^{-}\rangle=\frac{1}{\sqrt{2}}[|H_{1}V_{2}\rangle-|V_{1}H_{2}\rangle]$,
the correlation function $-\cos(2(\theta_{1}-\theta_{2}))$ is used.

\begin{figure}
\centering
\includegraphics[scale=0.35]{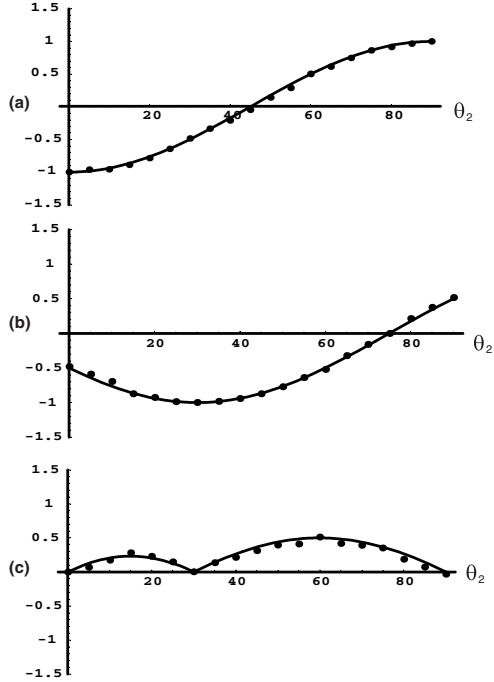}
\caption{\label{fig:4}(a) The measurement of correlation functions
,$C^{N}( a=0^{o},c=\theta_{2})$ and
(b)$C^{N}(b=30^{o}c=\theta_{2})$; (c) the plot of
$F(a,b,c)=|C^{N}(a=0^{o},b=30^{o})-C^{N}(a=0^{o},c)|-1-C^{N}(b=30^{o},c)$
showing $F(a,b,c) \not\leq 0$, maximum violation occurs at
$c=\theta_{2}=60^{o}$.}
\end{figure}

Maximum violation of Bell inequality of Eq.~\ref{eq:noise20} can be
demonstrated as analyzer A chooses polarization angles along the
axes a=$0^{o}$ and b=$30^{o}$ and analyzer B chooses along the axes
b=$30^{o}$ and c=$60^{o}$. First, we fixed the $a=\theta_{1}=0$,
then varied $c=\theta_{2}$ from $0^{o}$ to $90^{o}$ to obtain the
correlation function $C^{N}( a=0^{o},c=\theta_{2})$ as shown in
Fig.~\ref{fig:4}a. Second, we fixed $\theta_{1}=30^{o}$ and varied
$\theta_{2}$ from $0^{o}$ to $90^{o}$. The correlation function
$C^{N}(b=30^{o}, c=\theta_{2})$ is measured and shown in
Fig.~\ref{fig:4}b. By using the above measurements, we plot
$F(a,b,c)=|C^{N}(a=0^{o},b=30^{o})-C^{N}(a=0^{o},c)|-1-C^{N}(b=30^{o},c)$
as a function of $c=\theta_{2}$ as shown in Fig.~\ref{fig:4}c. The
solid lines in the figures are theoretical predictions by using
$C^{N}(\theta_{1},\theta_{2})=-\cos 2(\theta_{1}-\theta_{2})$. The
experimental results show that the maximum violation value is +0.5
occurs at the $c=\theta_{2}=60^{o}$, $F(a,b,c) \not\leq 0$.

The experiment has demonstrated nonlocality of two distant observers
based on superposition of one coherent light field and one coherent
noise field. This newly developed scheme can be implemented together
with the time-bin method for entanglement distributions and key
distributions, such as Ekert's protocol. For the prepared state
$|\psi^{-}\rangle
=\frac{1}{\sqrt(2)}[|H_{1}V_{2}\rangle-|V_{1}H_{2}\rangle]$, the
anti-correlation function is given by $-\cos(2(\theta_1 -
\theta_2))$, where $\theta_{1} = \theta_{2}$ for maximum
anti-correlation. When the beat signal at detector 1 has a positive
(negative) signal, the beat signal in detector 2 has a negative
(positive) signal. The random positive and negative beat signals can
be encoded for qubit implementation. The positive signal is encoded
to qubit "1" and the negative signal is encoded to qubit "0". This
encoding process can be conducted by using a comparator after the
detectors 1 and 2. If two coherent states, $|\alpha\rangle =
|\alpha|e^{-i\phi_{\alpha}}$ and $|\beta\rangle =
|\beta|e^{-i\phi_{\beta}}$, with low mean photon numbers are used,
their quantum phase fluctuations, $\phi_{\alpha}$, $\phi_{\beta}$
will play an essential role of randomness in this newly developed
scheme.

In conclusion, we have shown that the random and anti-correlated
beat signals at two spatially separated beams created by the
superposition of the coherent light field and the noise field can
exhibit the nonlocality and the duality properties of operators
$A_{1}$ and $B_{2}$. The scheme is perfect for long distance
entanglement distribution and key distribution. The experimental
observation has also implied that phase fluctuation and beam
splitter transformation are the origin creation of entanglement and
nonlocality for two coherent light fields.


\begin{acknowledgments}
The author would like to acknowledge that this work was done in
Department of Physics, Duke University under supervision of
Professor John Thomas. The author would also like to acknowledge
that this paper is prepared under the support of the start-up fund
from Department of Physics, Michigan Technological University
\end{acknowledgments}


\newcommand{\noopsort}[1]{} \newcommand{\printfirst}[2]{#1}
  \newcommand{\singleletter}[1]{#1} \newcommand{\switchargs}[2]{#2#1}

\end{document}